\documentclass[a4paper,12pt]{article}
\usepackage{a4wide}
\usepackage{graphicx}
\usepackage{refmergerus}
\usepackage{subfigure}
\begin{document}
\thispagestyle{headings}
\title{Study of the process $e^+ e^- \to \pi^+ \pi^- \pi^+ \pi^- \pi^0$ with CMD-2 detector}
\author{R.R.Akhmetshin,E.V.Anashkin,\and
V.M.Aulchenko, V.Sh.Banzarov, L.M.Barkov, S.E.Baru, \and
 A.E.Bondar, A.V.Bragin, D.V.Chernyak, S.I.Eidelman, N.S.Bashtovoy,  \and
G.V.Fedotovich, N.I.Gabyshev, A.A.Grebeniuk, D.N.Grigoriev, F.V.Ignatov, \and
P.M.Ivanov,  S.V. Karpov, V.F.Kazanin, B.I.Khazin, I.A.Koop, P.P.Krokovny,  \and
 L.M.Kurdadze, A.S.Kuzmin, I.B.Logashenko, P.A.Lukin, A.P.Lysenko,  \and 
K.Yu.Mikhailov, A.I.Milstein, I.N.Nesterenko, V.S.Okhapkin, \and
E.A.Perevedentsev, A.A.Polunin, A.S.Popov\thanks{contact person. e-mail:
Al.S.Popov@inp.nsk.su}, T.A.Purlatz, \and
  N.I.Root,  A.A.Ruban, N.M.Ryskulov,  A.G.Shamov, \and
Yu.M.Shatunov, B.A.Shwartz, A.L.Sibidanov,  \and
V.A.Sidorov, A.N.Skrinsky, V.P.Smakhtin, \and 
I.G.Snopkov,E.P.Solodov, P.Yu.Stepanov, \and
 A.I.Sukhanov,Yu.V.Yudin, S.G.Zverev \\ \vspace{2mm}
{\it Budker Institute of Nuclear Physics, Novosibirsk, 630090,
Russia} \\ \vspace{2mm}
J.A.Thompson \\ \vspace{2mm}
{\it University of Pittsburgh, Pittsburgh, PA, 15260, USA}
}
\date{}
\maketitle
\begin{abstract}
 The process $e^+ e^- \to \pi^+ \pi^- \pi^+ \pi^- \pi^0$
has been studied in the center of mass energy range 1280 -- 1380 MeV
using 3.0 pb$^{-1}$ of data collected with the CMD-2  detector 
in Novosibirsk. Analysis shows that the cross section of the five
pion production is dominated by the contributions of the $\eta\pi^+\pi^-$
and $\omega\pi^+\pi^-$ intermediate states.
\end{abstract}

 \section{Introduction}
 Measurements of the total cross section of $e^+ e^-$ annihilation into
hadrons at low energies as well as of the cross sections
of exclusive hadronic channels are of great interest for better
understanding the interactions of light quarks. This precise knowledge
is also necessary for calculations of effects such as
contributions of hadronic vacuum polarization to $(g-2)_\mu$ and
$\alpha(M^2_Z)$ \cite{ref:eid1}, tests of the Standard Model via the
hypothesis of conserved vector current (CVC) relating
 $e^+e^- \to $ hadrons to hadronic $\tau $-lepton decays
\cite{ref:tsai,ref:eid2}.
 \par  Annihilation of $e^+e^-$ into five pions  was first observed by
M3N \cite{ref:M3N}, DM1 \cite{ref:DM1_1,ref:DM1_2} and CMD
 \cite{ref:CMD} detectors. These  limited data samples
allowed one to  study this process qualitatively and estimate the
magnitude of the corresponding cross section. Later measurements by
various groups in Orsay and Novosibirsk (see \cite{ref:nd,ref:dm2,eta_dm2})
provided more detailed  information on the energy dependence of the
cross section for the processes $e^+e^-\to\eta\pi^+\pi^-$ and
$e^+e^-\to\omega\pi^+\pi^-$.
 \par Both isoscalar and isovector intermediate states can contribute to
the cross section of the reaction $e^+e^-\to \pi^+\pi^-\pi^+\pi^-\pi^0$.
In the isoscalar case the final $\omega(782)\pi^+\pi^-$ state is produced
via $\omega(782)$ and its excitations ($\omega(1420), \omega(1600)$),
whereas  the isovector contribution includes 
 the $\eta\pi^+\pi^- (\eta\rho(770))$
final state arising from the $\rho(770)$ and its excitations
($\rho(1450), \rho(1700)$). Since violation of isospin symmetry in
electromagnetic interactions is small, one can expect that the interference
between the modes with different isospin is also small, so that the two
cross sections could be found independently.

 \section{Experiment and event selection}
The experiment was performed with the CMD-2 detector at the $e^+e^-$ collider
VEPP-2M in Novosibirsk \cite {ref:aks}. The c.m. energy range
from the threshold of hadron production to 1.4 GeV was scanned with 
10 MeV steps. The highest energy attainable at VEPP-2M is rather  close to
the threshold of both final states studied, and therefore for the analysis of
five pion production about 3 pb$^{-1}$ of data collected in the energy range
1280 to 1380 MeV were used.

The tracking system of the detector consists of a drift chamber with about
250 $\mu$ resolution transverse to the beam  and a  proportional Z-chamber
used for the trigger, both inside a superconducting solenoid with a field
of 1 T. Photons are detected in the barrel CsI  calorimeter with  8-10\% energy
resolution and the endcap BGO calorimeter with  6\% energy resolution.
More details on the detector can be found elsewhere \cite{ref:aks}.
 \par To select events of the processes under study the following cuts
were applied:
  \begin{enumerate}
   \item Four charged tracks and two or more photons are found.
   \item The sum of the charges of all tracks is equal to zero.
   \item For each track the Z coordinate of the origin  is close to the
interaction point -- $|Z_{tr}|<7$ cm.
   \item The impact parameter of each track is small enough --- $R_{min} < 0.5$ cm.
  \end{enumerate}
 The background remaining after applying these cuts is mostly caused by the
 process  $ e^+e^- \to \pi^+\pi^-\pi^+\pi^- $  where extra photons appear
 because of the splitting of clusters from charged particles.
 \par The events selected by the above process were kinematically
 reconstructed  with the following constraints:
 \[ \sum_{i=1,4} E_{ch}(i) + \sum_{j=1,2} E_{\gamma}(j) = 2\cdot E_{beam}, \]
 \[ \sum_{i=1,4} \vec P_{ch}(i) + \sum_{j=1,2} \vec P_{\gamma}(j) = 0 ,\]
 where $E_{ch}$, $\vec P_{ch}$ are energies and momenta of charged particles,
 while 
$E_\gamma$ and $ \vec P_\gamma$ are those for photons.  
 The kinematic fit tries all possible photon pairs  in the event and the one
 with the smallest $\chi^2$ value is  used in further analysis. The
 events with thus chosen pairs of photons were again processed by kinematic
 reconstruction  with an additional constraint requiring the invariant
 mass of the two photons be equal to the $\pi^0$ mass:
 \[ m_{ \gamma \gamma} = m_{\pi^0}.\]
 The second stage helps  to increase the three pion
 invariant mass $m_{\pi^+\pi^-\pi^0}$ resolution.
 \par
\begin{figure}[htb]
  \centering\includegraphics[width=0.59\textwidth]{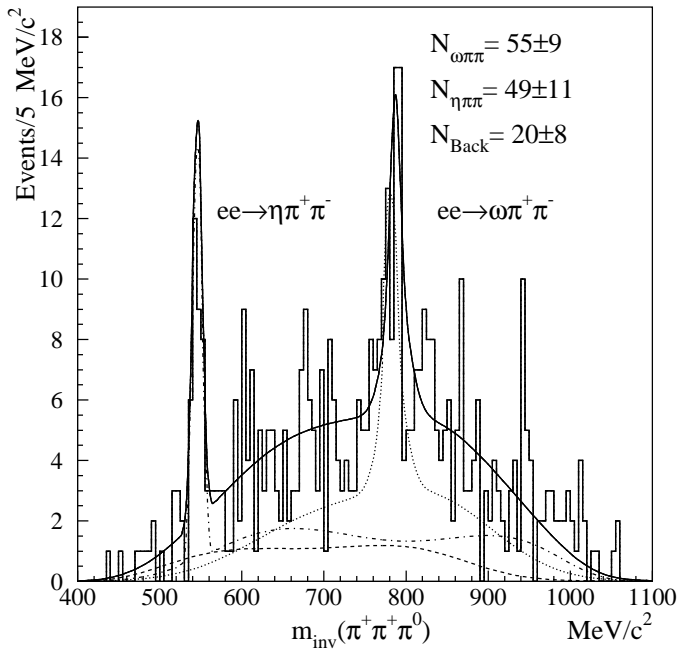}
   \caption {\label{tot_sp} The
    invariant mass $m_{\pi^+\pi^-\pi^0}$ for selected events. The solid curve
represents the  result of the fit, the dotted curve is the  contribution of
the $ \omega \pi^+ \pi^-$ mode, the dashed-dotted curve is that of 
    $ \eta \pi^+ \pi^-$, and the dashed curve shows the contribution of the
 background process     $ e^+ e^- \to \pi^+\pi^-\pi^+\pi^-$.}
 \end{figure}

 To further suppress  the background  from
 $\pi^+\pi^-\pi^+\pi^-$ events two additional cuts were used :
 \begin{enumerate}
 \item The energy of each of the two selected photons is  large enough ---
 $E_{\gamma}>40$ MeV. This cut efficiently rejects events with fake
 photons emerging because of the cluster splitting.
 \item The $\chi^2$ parameter characterizing the quality of the kinematic
fit is $\chi^2<4$.
 \end{enumerate}
 The $m_{\pi^+\pi^-\pi^0}$ distribution of selected events   is
 presented in Fig.~\ref{tot_sp}.
 \par
\begin{figure}
 \subfigure[\label{fm_om} $\,\,\,e^+e^- \to \omega \pi^+ \pi^-$ ]
{
   \includegraphics[width=0.45\textwidth]{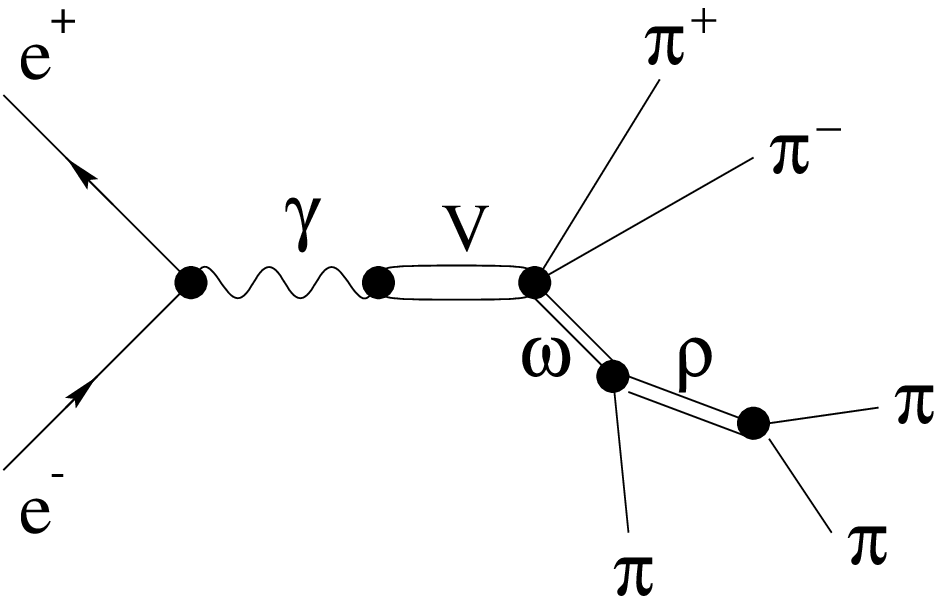}
}
 \hspace{0.02\textwidth}
   \subfigure[\label{fm_eta} $\,\,\,e^+e^- \to \eta \pi^+ \pi^-$ ]
{
   \includegraphics[width=0.45\textwidth]{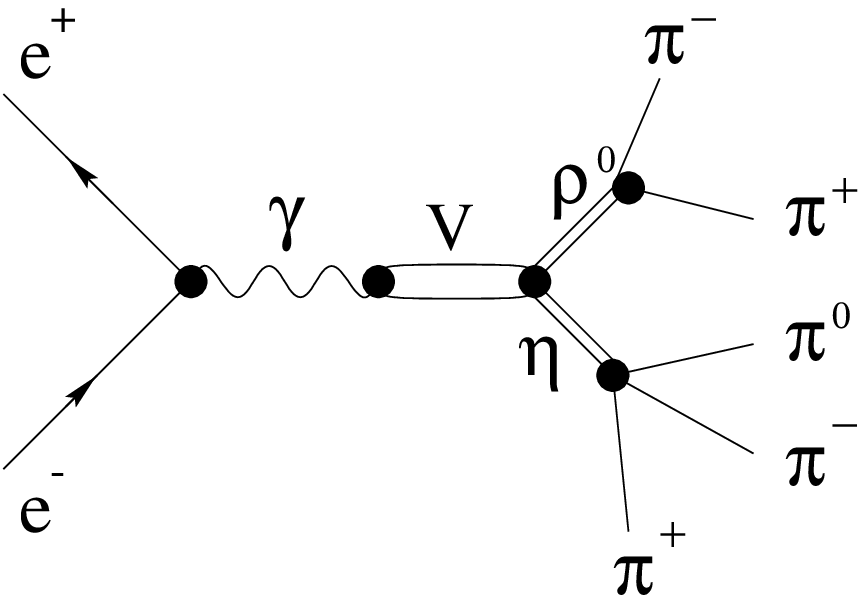}
}
 \caption{\label{fm} Feynman diagram for the processes under study.}
 \end{figure}

 Since all four possible combinations of pions  contribute,
a large combinatorial background is observed.
 The left peak in the figure corresponds to the $ \eta \pi^+ \pi^-$
 production whereas the right one is caused by the process
$ e^+ e^- \to \omega \pi^+ \pi^-$.
 The number of events for each process is found by fitting the histogram with
 the sum of three curves: $\eta \pi^+ \pi^-$, $\omega \pi^+ \pi^-$ and the 
 background of the
 $ e^+ e^- \to \pi^+\pi^-\pi^+\pi^-$ process.  The shape of the curves has
 been found from simulation based on  matrix elements
 consistent with the simplest Lorentz structure of the Feynman
 diagrams shown in Fig.~\ref{fm_om} and Fig.~\ref{fm_eta}.
 The interference between diagrams with different charges of intermediate
 $\rho$ mesons and with transpositions of identical  pions in the
 final state has been taken into account. Simulation of the background
 process $ e^+ e^- \to \pi^+ \pi^- \pi^+ \pi^- $ was performed according
 to Ref.~\cite{ref:nroot}. The procedure of statistical separation for events from different 
 processes was checked by using simulation and the systematic  error associated with this 
 procedure was found to be negligible. 

 \section{Results}
 The detection efficiency for our selection criteria was found to be
 25\% and 23\% for  $\omega\pi^+\pi^-$ and $\eta\pi^+\pi^-$
respectively, with no significant energy dependence. The radiative 
corrections for each process were calculated
following Ref.~\cite{ref:fad}.

\begin{figure}[htb]
  \subfigure[\label{om_sec}  $e^+e^-\to\omega\pi^+\pi^-$ cross section.]	
{ 
    \begin{minipage}[t]{0.49\textwidth}
     \centering\includegraphics[width=0.99\textwidth]{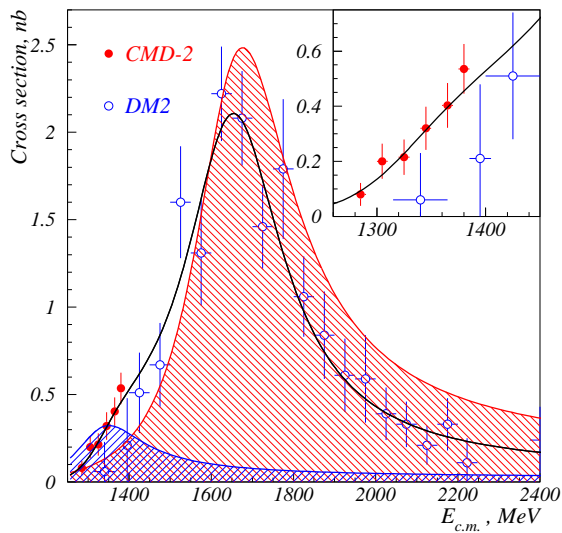}
        \end{minipage}
}
  \subfigure[\label{eta_sec}  $e^+e^-\to\eta\pi^+\pi^-$ cross section.]	
{
    \begin{minipage}[t]{0.49\textwidth}
     \centering\includegraphics[width=0.99\textwidth]{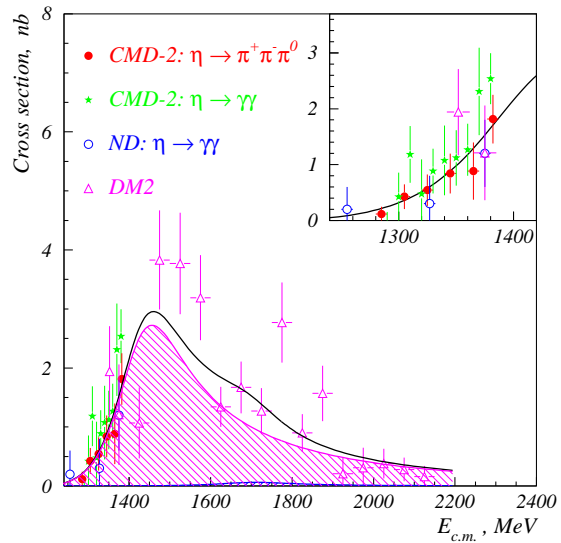}
        \end{minipage}
}
\caption{\label{cr_sec}Energy dependence of the cross sections for the 
processes under study: 
(a) - $e^+e^- \to \omega\pi^+\pi^-$, (b) - $e^+e^- \to \eta\pi^+\pi^-$.
In each Figure individual contributions of two higher resonances are shown by
hatched areas. The inserts show the energy range studied by CMD-2 in 
larger scale. 
} 
 \end{figure}

The value of the radiative correction varied with energy from 
-16.5\% to -20.0\% for each process.
The number of signal events and  corresponding cross sections
are  presented in  Table \ref{odin} for each energy and shown in
Figs. \ref{om_sec} and \ref{eta_sec} in comparison with the results of other
experiments~\cite{ref:nd,ref:dm2,eta_dm2,ref:dip}.
The systematic uncertainty  mainly comes from  the difference in $\chi^2$ distributions for 
simulation and experimental data  and was estimated as $15$\%.
Our results for the cross section of the process
$e^+e^- \to \omega\pi^+\pi^-$ are higher than those from
DM2, but they are not incompatible within the large statistical and
systematic uncertainties.  The cross section of the process
$e^+e^- \to \eta\pi^+\pi^-$ agrees with the results of
other measurements by ND and DM2 as well as with the
preliminary results from CMD-2 \cite{ref:dip} obtained using the
decay mode $\eta \to \gamma \gamma$.

 The number of background events found in this way is consistent
with the estimate based on the value of the cross section for
the process $e^+e^-\to\pi^+\pi^-\pi^+\pi^-$ obtained by CMD-2
\cite{ref:nroot}. One can therefore conclude that the cross section
of the five pion production is saturated by the $\omega\pi^+\pi^-$ and
$\eta\pi^+\pi^-$ intermediate states.

 To understand whether the data 
allow a small contribution from the $\pi^+\pi^-\pi^+\pi^-\pi^0$ final state 
other than 
$\omega\pi^+\pi^-$ or $\eta\pi^+\pi^-$, the following analysis was 
carried out. A strict cut on 
$\chi^2 < 1$ was imposed considerably suppressing the background from 
$\pi^+\pi^-\pi^+\pi^-$. The expected shape of the $m_{\pi^+\pi^-\pi^0}$
distribution was determined from a simulation  of the process 
$e^+e^-\to\pi^+\pi^-\pi^+\pi^-\pi^0$ assuming a constant matrix element. 
Next  a fit of the experimental distribution of  
$m_{\pi^+\pi^-\pi^0}$ was performed assuming all four contributions discussed 
above. 
Results of this fit 
in terms of the upper limits on the possible contribution to the cross section of 
$\pi^+\pi^-\pi^+\pi^-\pi^0$
production at 95\% confidence level are shown in the last column of Table \ref{odin}.     
One can  see that these limits are typically much smaller than the
$\omega\pi^+\pi^-$ cross section indicating  the smallness of other
intermediate mechanisms of $\pi^+\pi^-\pi^+\pi^-\pi^0$ production. 

Results of the fit are shown in Figs.~\ref{om_sec}, \ref{eta_sec} and presented in Table \ref{dwa}. The  
$\chi^2$/n.d.f. value is equal to 15.1/19 and 37/31 for  $\omega\pi^+\pi^-$ and 
$\eta\pi^+\pi^-$ respectively. Hatched areas show separate contributions 
for each resonance. One can see that the 
contributions of $\omega(1420)$ for 
$\omega\pi^+\pi^-$ and $\rho(1700)$  for $\eta\pi^+\pi^-$ are negligible compared to the  
$\omega(1600)$ and $\rho(1450)$ respectively and are not  statistically significant. 
The fit with one resonance only was additionally performed which results in    
$\chi^2$/n.d.f = 23/23 and $\chi^2$/n.d.f = 43/35 for $\omega\pi^+\pi^-$ and $\eta\pi+\pi^-$ respectively.
\par To estimate the possible variation of parameters we performed the fit of 
the $\omega\pi^+\pi^-$ cross section  under the assumption that  the total width is completely determined by the decay
$\omega_{1,2}\to\rho\pi\to\pi^+\pi^-\pi^0$. In accordance with 
\cite{ref:PDG}, the branching ratio of the 
$\omega(1420)$ is dominated by the $\rho \pi$ decay while for the 
$\omega(1600)$ it is about $\sim$ 50 \%, so that 
this assumption does not look completely unreasonable. In this case 
$\Gamma_{1,2}(s) = \Gamma^0_{1,2}\cdot\frac{W_{\rho\pi}(s)}{W_{\rho\pi}(m_{1,2})}$, where
$W_{\rho\pi}$ is the squared matrix element of the $\rho\pi$ decay integrated over the phase space.
$\Gamma^0_{1,2}$ as well as  $m_{1,2}$, $\phi$ and $\sigma_{1,2}$  are 
parameters of the fit. $B_{e^+e^-}\,\,B_{fin}$ 
 presented in Tables~\ref{dwa} and \ref{tri}  were calculated  in accordance with the relation 
$B_{e^+e^-}\,\,B_{fin} = \frac{\sigma(m^2)}{12\,\pi}\,\,m^2$. 

The fit results of which are presented in Table~\ref{tri} 
has $\chi^2$/n.d.f = 13/19 indicating a good fit to this hypothesis. 
As one can conclude from comparison of Table~\ref{dwa} and Table~\ref{tri},
the parameters of the $\omega(1420)$ are more or less stable while for the $\omega(1600)$ the level of their variation 
is of the order of two standard deviations.  
More precise experimental information collected in the whole energy range from 1.4 to 2 GeV
is obviously necessary to determine unambiguously the   properties of the 
$\rho$(770) and $\omega$(782) excitations.

\begin{table}
\begin{center}
 \begin{tabular}{|c|c|c|c|c|c|c|}
  \hline
   \multicolumn{1}{|c|}{}&\multicolumn{1}{|c|}{}&\multicolumn{2}{|c|}
  {$\omega\pi\pi$}&\multicolumn{2}{|c|}{$\eta\pi\pi$}&\multicolumn{1}{|c|}{ 5$\pi$}\\ \hline
 $E_{c.m.}$,     & L, nb$^{-1}$          & $n_{eff}$    & $\sigma$, nb & $n_{eff}$ & $\sigma$, nb & $\sigma$, nb \\ 
  MeV            &                       &              &              &           &              & 95\% CL      \\ \hline
 1285    & 536   & 9.6  $\pm$ 6.8 & 0.07 $\pm$ 0.05 & 3.2  $\pm 3.7 $  & 0.11 $\pm$ 0.13&0.06 \\ \hline
 1305    & 510   & 15.5 $\pm$ 4.7 & 0.20 $\pm$ 0.06 & 5.5  $\pm 2.8 $  & 0.42 $\pm$ 0.22&0.12 \\ \hline
 1325    & 556   & 32.6 $\pm$ 9.2 & 0.21 $\pm$ 0.06 & 13.2 $ \pm 6.8 $ & 0.54 $\pm$ 0.28&0.17 \\ \hline
 1345    & 440   & 33.2 $\pm$ 8.3 & 0.32 $\pm$ 0.08 & 14.5 $\pm 6.0 $ & 0.84 $\pm$ 0.35 &0.10\\ \hline
 1365    & 415   & 30.5 $\pm$ 6.1 & 0.40 $\pm$ 0.08 & 13.5 $\pm 7.8 $ & 0.88 $\pm$ 0.51 &0.11\\ \hline
 1380    & 541   & 55.4 $\pm$ 9.0 & 0.53 $\pm$ 0.09 & 49.0 $\pm 11.4$ & 1.81 $\pm$ 0.43 &0.19\\ \hline
 Total   & 2998  & 176.8$\pm$ 18.4& ---             & 98.9 $\pm17.2$ & ---              & ---\\ \hline
 \end{tabular}
\caption{}
\label{odin}
\end{center}
\end{table}

\begin{table}[tbh]
\begin{center}
 \begin{tabular}{|c|c|c|c|c|}
  \hline
                & $\omega(1420)$ & $\omega(1600)$ &$\rho(1450)$   & $\rho(1700)    $\\ \hline    
 $\sigma(m^2)$,nb & $0.31\pm0.27$  & $2.41\pm0.33$  & $2.53\pm0.58$ & $ 0.29\pm0.24  $\\ \hline
 $m$, Mev       & $1329\pm64  $  & $1652\pm35$    & $1421\pm15  $ &  ---  \\ \hline
 $\Gamma$, MeV  & $198 \pm73  $  & $284 \pm31$    & $211 \pm31  $ &  ---  \\ \hline
 $\phi$, rad    &    0          & $3.3\pm0.3$  &      0        & $ 1.9\pm0.5  $\\ \hline
 $B_{e^+e^-}\,\,B_{fin}\cdot10^{-7}$ &$0.37\pm0.32$&$4.5\pm0.6$ &$3.5\pm0.8$&$0.6\pm0.5$\\\hline
 \end{tabular}
\end{center}
\caption{Parameters from the fit by two resonances from Figs.~\ref{om_sec},\ref{eta_sec}}
\label{dwa}
\end{table}

\begin{table}[ht]
\begin{center}
 \begin{tabular}{|c|c|c|}
  \hline
                & $\omega(1420)$ & $\omega(1600)$  \\ \hline    
 $\sigma(m^2)$, nb & $0.10\pm 0.10$ & $1.82 \pm 0.32$ \\ \hline
 $m$, MeV       & $1373\pm 70  $ & $1705 \pm 26  $ \\ \hline
 $\Gamma$, MeV  & $188 \pm45   $ & $370  \pm 25  $ \\ \hline
 $\phi$, rad    &     0          & $2.3 \pm 1.0$   \\ \hline
 $B_{e^+e^-}\,\,B_{fin} $ &$(0.13\pm0.13)\,\,10^{-7}$&$(0.36\pm0.06)\,\,10^{-6}$ \\\hline
 \end{tabular}
\end{center}
\caption{Parameters from the fit of the $\omega\pi^+\pi^-$ cross section by two resonances with the energy
dependent width.}
\label{tri}
\end{table}

\section{Conclusions}
The cross section of $e^+e^-$ annihilation into five pions
has been measured in the energy range 1.28 - 1.38 GeV using
the CMD-2 detector. The cross section is saturated by
the contributions of $\omega\pi^+\pi^-$ and $\eta\pi^+\pi^-$
intermediate mechanisms. The energy dependence of the cross
sections is compatible with the generally accepted pattern
of two higher mass excitations for both $\rho$(770) and $\omega$(782) mesons, 
although an acceptable fit was obtained with one isoscalar and one 
isovector resonace only ($\omega(1600)$ and $\rho(1450)$). Parameters of 
the resonances are rather  model dependent and more detailed information 
at c.m. energies above 1.4 GeV is required for their unambiguous determination.

 \end{document}